\documentclass[12pt,a4paper]{article}
\usepackage{amssymb,epsfig,amsmath,latexsym}
\newtheorem{theorem}{Theorem}[section]
\newtheorem{proposition}[theorem]{Proposition}
\newtheorem{definition}[theorem]{Definition}

\newtheorem{corollary}[theorem]{Corollary}
\newtheorem{example}[theorem]{Example}
\newcommand{\pf}{\noindent{\textit{Proof:} }}
\newcommand\Q[1]{$#1$ \quad}

\newcommand\abs[1]{\lvert#1\rvert}

\newcommand{\ZZ}{\mathbb{Z}}
\newcommand{\FF}{\mathbb{F}}

\newcommand{\R}{\mathcal{R}}
\newcommand{\C}{\mathcal{C}}
\begin{document}
\title{One-Sided $k$-Orthogonal Matrices \\ Over Finite Semi-Local Rings \\ And Their Codes } 

\author{Virgilio P. Sison\\Institute of Mathematical Sciences and Physics\\University of the Philippines Los Ba\~{n}os\\College 4031, Laguna, Philippines\\Email: {\tt vpsison@up.edu.ph} \\ \\ Charles R. Repizo\\Department of Mathematics and Physics\\ University of Santo Tomas\\ Espa\~{n}a Blvd., Sampaloc, Manila 1008, Philippines\\ Email: {\tt crrepizo@ust.edu.ph} }

\maketitle

\begin{abstract}
Let $R$ be a finite commutative ring with unity $1_R$ and $k \in R$. Properties of one-sided $k$-orthogonal $n \times n$ matrices over $R$ are presented. When $k$ is idempotent, these matrices form a semigroup structure. Consequently new families of matrix semigroups over certain finite semi-local rings are constructed. When $k=1_R$, the classical orthogonal group of degree $n$ is obtained. It is proved that, if $R$ is a semi-local ring, then these semigroups are isomorphic to a finite product of $k$-orthogonal semigroups over fields. Finally, the antiorthogonal and self-orthogonal matrices that give rise to leading-systematic self-dual or weakly self-dual linear codes are discussed.
\end{abstract}

\section{Introduction}
This study extends the classical theory of orthogonal matrices over fields to the so-called one-sided (left or right) $k$-orthogonal matrices over rings, where $k$ is some fixed scalar element. This new class of one-sided $k$-orthogonal matrices is examined in the context of finite commutative rings with unity. In particular, we focus our attention to finite semi-local rings with at least two maximal ideals in order to fully utilize the Chinese Remainder Theorem. Various properties of $k$-orthogonal matrices are described and applications of these matrices in constructing linear codes are explored. It is important to note at this point that we have given a different definition of $k$-orthogonal matrices which may vary from those found in other literature.

The material is organized as follows: Section~\ref{sect:prelim} provides the important conceptual framework while Section~\ref{sect:result} contains the main results. MAGMA${}^{\circledR}$ routines are designed to generate new examples of one-sided and two-sided $k$-orthogonal matrices and to construct the codes from these matrices. A simple algorithm to generate the set of idempotent elements of an arbitrary finite ring is also written. Section~\ref{sect:recom} gives suggestions for further research.

\section{Preliminaries and definitions}
\label{sect:prelim}

This section tackles briefly the concept of orthogonal matrices and the general linear group. The classical theory of orthogonal matrices over the finite field $\FF_q$ is fully discussed by MacWilliams \cite{macw}. Some important ring-theoretic concepts, such as coprime and maximal ideals, idempotents, local rings, semi-local rings and finite chain rings, are also introduced here. For a more thorough treatment of these topics, the reader is referred to \cite{han, hun, macd}.  

\subsection{Ring of matrices and the orthogonal group}

Let $R$ be a commutative ring with unity $1_R \ne 0$ and $M_n(R)$ the unital noncommutative ring of square $n \times n$ matrices over $R$. A matrix $A \in M_n(R)$ is {\it nonsingular} or {\it invertible} if and only if $\det(A)$ is a unit in $R$, or equivalently, there exists $B \in M_n(R)$ such that $AB = BA = I_n$, where $I_n$ is the identity matrix. The set of invertible matrices in $M_n(R)$ forms a noncommutative multiplicative group and is denoted by $GL_n(R)$. A matrix $A$ is {\it symmetric} if $A = A^T$, where $A^T$ is the transpose of $A$. It is called {\it orthogonal} provided the following holds.
\begin{equation} \label{ortho} A^TA=AA^T=I_n.\end{equation} In other words, an orthogonal matrix is an invertible matrix whose unique two-sided inverse is its transpose. In fact the set of orthogonal matrices over $R$, denoted $O_n(R)$, is a (not necessarily normal) subgroup of $GL_n(R)$, and is called the {\it orthogonal group of degree $n$}. It should be remarked that, if $A \in O_n(R)$, then $\det(A)$ is either the identity or an involution, but not conversely.

Let $A=[a_{ij}]$, $B=[b_{ij}]\in M_n(R)$, and $I$ an ideal of $R$. Then $A$ is said to be {\it congruent} to $B$ modulo $I$, written as $A\equiv_n B\pmod{I}$, if and only if $a_{ij}\equiv b_{ij}\pmod{I}$, that is $a_{ij}-b_{ij}\in I$, for all $i,j=1,2,...,n$. This is simply component-wise application of the usual additive subgroup relation on an ideal. It is shown in \cite{han} that $\equiv_n$ is, quite expectedly, a congruence relation in $M_n(R)$.  

Two ideals $I$ and $J$ of $R$ are said to be {\it coprime} if $I+J=R$. A set of nontrivial ideals $\mathcal I = \{I_1,I_2,...,I_m\}$ in $R$ is {\it pairwise coprime} if $I_j+I_k=R$ for all $j,k=1,2,...,m$ with $j\neq k$. Note that, if $\mathcal I$ consists of distinct maximal ideals, then necessarily $\mathcal I$ is pairwise coprime. The Chinese Remainder Theorem gives the following powerful result for matrix rings.

\begin{theorem} [Han, 2006] \label{han2006} Let $m$ and $n$ be any positive integers and $I_1,$ $I_2,$...,$I_m$ be ideals in a ring $R$.Then there is a monomorphism of rings $$\theta:M_n\bigl(R/I_1\cap I_2\cap...\cap I_m\bigr)\rightarrow M_n(R/I_1)\times M_n(R/I_2)\times \ldots\times M_n(R/I_m)$$ If $I_1,I_2,\dots,I_m$ are pairwise coprime, then $\theta$ is an isomorphism. 
\end{theorem}
This result extends naturally to the general linear group $GL_n(R)$, and when $R$ is finite and the ideals $I_j$ are maximal, it gives a convenient formula for $|GL_n(R)|$ as a product of the $m$ factors $|GL_n(\FF_{q_j})|$, in terms of the residue fields $\FF_{q_j}$, given by
\begin{equation} \label{general}
|GL_n(\FF_{q_j})|=q_j^{\tfrac{n(n-1)}{2}}\displaystyle\prod_{i=1}^{n}(q_j^i-1), j=1,2,\ldots,m
\end{equation}

\subsection{Finite semi-local rings}

A {\it semi-local ring} is a commutative ring $R$ with unity with a finite number of maximal ideals. In particular, if $R$ has a unique maximal ideal, then $R$ is said to be a {\it local ring}. Hence a local ring is semi-local but not conversely. The quaternary ring $\FF_2+v\FF_2$, where $v^2=v$, is a semi-local ring with two maximal coprime ideals $(v)$ and $(1+v)$. The ring $\ZZ_6$ of integers modulo $6$ is also a semi-local ring with two maximal coprime ideals $(2)$ and $(3)$. A finite commutative ring with unity whose ideals are linearly ordered by inclusion is a {\it finite chain ring}. The ring $\ZZ_{p^r}$ of integers modulo $p^r$, $p$ prime and $r \ge 1$ an integer, is a local ring with unique maximal ideal $(p)$, and is a finite chain ring of length $r$. 

An {\it idempotent} of a ring is an element $a$ with the property that $a^2=a$, that is, its square is itself. A {\it Boolean ring} is a ring in which all its elements are idempotent. The ring $\FF_2+v\FF_2$, where $v^2=v$, is an example of a Boolean ring. It is also known that, if $a_1,a_2,..,a_m$ are idempotent elements of rings $R_1,R_2,...,R_m$, respectively, then the $m$-tuple $(a_1,a_2,...,a_m)$ is idempotent in the cross product $R_1\times R_2\times...\times R_m$. The integer ring $\ZZ_n$ has $2^k$ idempotent elements where $k$ is the number of distinct prime factors of $n$. Specifically, the ring $\ZZ_6$ is not Boolean but it has $4$ idempotents out of $6$ elements, namely, $\{0,1,3,4\}$. Finite local rings with unity contain no idempotents except the trivial elements $0$ and $1$. 

\subsection{Orthogonal codes}

Let $k < n$ be positive integers. A rate-$k/n$ {\it linear code} $\C$ of length $n$ over $R$ is an $R$-submodule of $R^n$ given by the set $\C=\{v\in R^n\;|\;v=uG, u\in R^k\}$, where $G$ is a $k \times n$ matrix whose rows span the elements of $\C$ which are called {\it codewords}. If no proper subset of the rows of $G$ generates $\C$, then $G$ is called a {\it generator matrix} of $\C$. The code $\C$ is a {\it leading-systematic code} if it has a systematic generator matrix of the form $G=[I_k:A]$, where $I_k$ is the $k \times k$ identity matrix and $A$ is a $k \times (n-k)$ matrix, in which case, $\C$ is a {\it free code} since the rows of $G$ form a basis for $\C$. We recall that $\C$ is a {\it weakly self-dual code} if $\C \subseteq \C^{\perp}$, where $\C^{\perp}=\{u\in R^n\;|\;\langle u,v\rangle=0,\;\text{for all}\;v\in\C\}$ is the orthogonal or dual code of $\C$, and that $\C$ is {\it self-dual} if $\C = \C^{\perp}$.

Massey in \cite{mas1} gives specific notions of orthogonal matrices over fields depending on the result of the product of a given matrix with its transpose. An $n \times n$ matrix $O$ is an {\it antiorthogonal matrix} if $O^TO=-I_n$, and is a {\it self-orthogonal matrix} if $O^TO=0_n$, where $0_n$ is the all-zero matrix. Orthogonal, antiorthogonal and self-orthogonal matrices are then used to construct certain linear codes over fields. It is shown in \cite{mas1} that a leading-systematic linear code is self-dual if and only if the matrix $A$ in its generator matrix $G=[I:A]$ is an antiorthogonal matrix. The conditions for the existence of antiorthogonal matrices over a Galois field are stated in \cite{mas2}.  

A {\it linear complementary dual code} $\C$ (or LCD code) is a code satisfying $\C\cap\C^{\perp}=\{\mathbf{0}\}$. Given the generator matrix $G$ of a code $\C$, then $\C$ is self-dual if and only if $GG^T$ is a nonsingular matrix, as shown in \cite{mas3} for the field case. Moreover, a leading systematic linear code with $G=[I:A]$ is an LCD code if $A$ is a row self-orthogonal matrix, or equivalently, if $G$ is a row-orthogonal matrix.

\section{Results and discussion}
\label{sect:result}

We extend the theory of orthogonal matrices by considering left or right sided $k$-orthogonal matrices over a finite commutative ring $R$ with unity, where $k$ is some fixed element of $R$. We begin with the structure of one-sided $k$-orthogonal matrices in $M_n(R)$.

\subsection{$k$-orthogonal matrices}

\begin{definition} \label{korth} Let $k$ be a fixed element of $R$. A matrix $A\in M_n(R)$ is said to be left [resp. right] $k$-orthogonal if and only if $A^TA=kI_n$ [resp. $AA^T=kI_n$], where $I_n$ is the identity of $M_n(R)$. A matrix $A$ is said to be two-sided $k$-orthogonal (or simply, $k$-orthogonal) if it is both left $k$-orthogonal and right $k$-orthogonal.
\end{definition}

\begin{example} \label{firstex} Given the matrix $A=\begin{bmatrix} 2 & 5 \\ 1 & 2 \end{bmatrix}\in M_2(\ZZ_6)$. Then $A^TA= AA^T = \begin{bmatrix} 5 & 0 \\ 0 & 5 \end{bmatrix}$ so that $A$ is a two-sided $5$-orthogonal matrix.
\end{example}
Obviously the transpose of a left [resp. right] $k$-orthogonal matrix is right [resp. left] $k$-orthogonal. Further, if $A$ is left [resp. right] $k$-orthogonal, then $[\det(A)]^2=k^n$, and specifically, if $R$ is Boolean, $\det(A)=k$.  However, a left [resp. right] $k$-orthogonal matrix is not necessarily right [resp. left] $k$-orthogonal, as given in the following example.

\begin{example}
Throughout this paper we will denote by $\R_2$ the Boolean ring $\FF_2+v\FF_2=\{0,1,v,1+v\}$, where $v^2=v$. Given $A=\begin{bmatrix} v & 0 \\ 1+v & 1 \end{bmatrix}\in M_2(\R_2)$, then $AA^T=\begin{bmatrix} v & 0 \\ 0 & v \end{bmatrix}$ so that $A$ is right $v$-orthogonal. The determinant of $A$ is $v$, that is, $A$ is not a unit in $M_2(\R_2)$. Furthermore, $A^TA=\begin{bmatrix} 1 & 1+v \\ 1+v & 1 \end{bmatrix}$, thus $A$ is not left $v$-orthogonal.
\end{example} 

\begin{example}
The matrix $A=\begin{bmatrix} 1+v & 1 \\ 1  & 1+v \end{bmatrix}\in M_2(\R_2)$ satisfies $A^TA=AA^T=vI_2$ and hence is two-sided $v$-orthogonal.
\end{example}
\begin{example} \label{selforth} The matrix $B=\begin{bmatrix} 1+v & 0 & 1+v \\ 1 & v & 1+v \\ v & v & 0 \end{bmatrix}\in M_3(\R_2)$ satisfies $B^TB$ $=BB^T=0_3$, where $0_3$ is the $3 \times 3$ all-zero matrix. Therefore, $B$ is a two-sided $0$-orthogonal matrix.
\end{example}

It is important to note that, if $k \in R$ is idempotent, the scalar matrix $kI_n$ is both left $k$-orthogonal and right $k$-orthogonal. The statement is false when $k$ is not idempotent. For instance, the matrix $2I_n \in M_n(\ZZ_6)$. One can easily verify that this matrix is neither left $2$-orthogonal nor right $2$-orthogonal. We extend this statement in the proposition below. 

\begin{proposition} \label{prop1}
Let $k$ be an idempotent element. If $A$ is left [resp. right] $k$-orthogonal, then $kA$ is also left [resp. right] $k$-orthogonal, but not conversely.
\end{proposition}
\pf Quite easily, $(kA)^T(kA)=k^{2n}A^TA=k^{2n+1}I_n=kI_n.$ \Q{\Box}

\begin{example} 
The matrix $A=\begin{bmatrix} 0 & 1+v \\ 1 & v \end{bmatrix}$ is right $(1+v)$-orthogonal. Then $(1+v)A =\begin{bmatrix} 0 & 1+v \\ 1+v & 0 \end{bmatrix}$ is also right $(1+v)$-orthogonal. However, the reversal matrix $\begin{bmatrix} 0 & 1 \\ 1 & 0 \end{bmatrix}$ is not right $(1+v)$-orthogonal. 
\end{example}

We then construct the set of left $k$-orthogonal matrices and the set of right $k$-orthogonal matrices, and denote these sets by $LO_n(k,R)$ and $RO_n(k,R)$, respectively. When $k$ is idempotent, these sets are multiplicative semigroups.  

\begin{proposition} Let $k$ be an idempotent element of $R$, and $n$ be a positive integer, $n\geq2$. The sets $LO_n(k,R)$ and $RO_n(k,R)$ of left and right $k$-orthogonal matrices in $M_n(R)$, respectively, are semigroups under matrix multiplication.
\end{proposition}
\pf  The set $LO_n(k,R)=\{A\in M_n(R)\;|\;A^TA=kI_n\}$ is nonempty because the scalar matrix $kI_n$ is an element. Since $LO_n(k,R)\subseteq M_n(R)$, associativity is inherent. We just need to show that $LO_n(k,R)$ is closed under matrix multiplication. Let $A,B\in LO_n(k,R).$ Then $(AB)^T(AB)$ $=(B^TA^T)(AB)$ $=B^T(kI_n)B$ $=k(B^TB)$ $=k(kI_n)$$=k^2I_n$ $=kI_n$. The proof for $RO_n(k,R)$ is analogous. \Q{\Box}
\vskip .1in
The identity $I_n$ of $M_n(R)$ is not in $LO_n(k,R)$ and in $RO_n(k,R)$ in general and hence these semigroups are not monoids. The set of two-sided $k$-orthogonal matrices, denoted $O_n(k,R)$, satisfies by definition, \begin{equation} \label{twosided} O_n(k,R) = LO_n(k,R) \cap RO_n(k,R)\end{equation} and therefore is a nonempty multiplicative semigroup as well, as $kI_n \in O_n(k,R)$. The semigroup $O_n(1_{R},R)$ is actually the orthogonal group $O_n(R)$.

\begin{example} \label{exa16}
The orthogonal group $O_2(1,\ZZ_6)$ contains exactly 16 two-sided $1$-orthogonal matrices $A$ with determinant $1$ or $5$, the only units of $\ZZ_6$, so that $[det(A)]^2=1$. $O_2(1,\ZZ_6)$ is a (not normal) subgroup of $GL_2(\ZZ_6)$ with 288 elements. 
\end{example}

\begin{proposition} Let $k$ and $k'$ be idempotent elements of $R$, and $n \ge 2$ an integer. Then, either $LO_n(k,R)\cap LO_n(k',R)=\varnothing$ or $LO_n(k,R)=LO_n(k',R).$ Similarly for the right $k$-orthogonal semigroups.
\end{proposition}
\pf Suppose $LO_n(k,R)\cap LO_n(k',R) \ne \varnothing$. Then there is a matrix $A$ such that $A\in LO_n(k,R)\cap LO_n(k',R)$. It follows that $A^TA=kI_n$ and $A^TA=k'I_n$ and thus $kI_n=k'I_n$ so that $k=k'$. \Q{\Box} 
\begin{example}
The left orthogonal semigroup $LO_2(1+v,\R_2)$ consists of the matrices $$\begin{bmatrix} 1+v & 0 \\ 0 & 1+v \end{bmatrix},
\begin{bmatrix} 1 & v \\ v & 1 \end{bmatrix},
\begin{bmatrix} 0 & 1+v \\ 1+v & 0 \end{bmatrix},
\begin{bmatrix} v & 1 \\ 1 & v \end{bmatrix},$$
$$\begin{bmatrix} 1 & 0 \\ v & 1+v \end{bmatrix},
\begin{bmatrix} v & 1+v \\ 1 & 0 \end{bmatrix},
\begin{bmatrix} 0 & 1 \\ 1+v & v \end{bmatrix},
\begin{bmatrix} 1+v & v \\ 0 & 1 \end{bmatrix}.$$
\noindent You will see in Table \ref{table:tab1} that $LO_2(1+v,\R_2)\cap LO_2(v,\R_2)=\varnothing$.
\end{example}

We use the fact that the transpose of a left [resp. right] $k$-orthogonal matrix is right [resp. left] $k$-orthogonal to claim the following theorem.
\begin{theorem} \label{card}
For any idempotent element $k \in R$ and positive integer $n \ge 2$, we have $\abs{LO_n(k,R)}=\abs{RO_n(k,R)}$.
\end{theorem}    
\pf The map $f:LO_n(k,R)\longrightarrow RO_n(k,R)$ defined by \begin{equation}\label{transpose} A\mapsto A^T \end{equation} is a bijection. \Q{\Box}
\vskip .1in
\noindent This simple correspondence implies that $LO_n(k,R) \neq RO_n(k,R)$, even if $R$ is a field. In Section \ref{subsec:iso}, we shall give the formulas for these cardinalities when $R$ is finite. 
\vskip .1in
Table \ref{table:tab1} illustrates Theorem~\ref{card} with one-sided orthogonal semigroups $LO_2(v,\R_2$) and $RO_2(v,\R_2)$ consisting of eight matrices each via the correspondence in (\ref{transpose}). Since $\R_2$ is Boolean, each left [resp. right] $v$-orthogonal matrix has determinant $v$. The matrices in the first four rows are the elements of the two-sided orthogonal semigroup $O_2(v,\R_2)$ which are circulant symmetric. We prove this observation in a more general sense in Proposition~\ref{abk}. 
\begin{table}[ht]
\renewcommand{\arraystretch}{1.3}
\label{table:tab1}
\caption{Left and right $v$-orthogonal semigroups over $\R_2=\FF_2+v\FF_2$ }
\centering
\begin{tabular}{c|c}\hline \hline
$LO_2(v,\R_2)$ & $RO_2(v,\R_2)$ \\ \hline 
$\begin{bmatrix} v & 0 \\ 0 & v \end{bmatrix}$ & $\begin{bmatrix} v & 0 \\ 0 & v \end{bmatrix}$ \\
\hline
$\begin{bmatrix} 1 & 1+v \\ 1+v & 1 \end{bmatrix}$ & $\begin{bmatrix} 1 & 1+v \\ 1+v & 1 \end{bmatrix}$ \\
\hline
$\begin{bmatrix} 0 & v \\ v & 0 \end{bmatrix}$ & $\begin{bmatrix} 0 & v \\ v & 0 \end{bmatrix}$ \\
\hline
$\begin{bmatrix} 1+v & 1 \\ 1 & 1+v \end{bmatrix}$ & $\begin{bmatrix} 1+v & 1 \\ 1 & 1+v \end{bmatrix}$ \\
\hline
$\begin{bmatrix} 1 & 0 \\ 1+v & v \end{bmatrix}$ & $\begin{bmatrix} 1 & 1+v \\ 0 & v \end{bmatrix}$ \\
\hline
$\begin{bmatrix} v & 1+v \\ 0 & 1 \end{bmatrix}$ & $\begin{bmatrix} v & 0 \\ 1+v & 1 \end{bmatrix}$ \\
\hline
$\begin{bmatrix} 0 & 1 \\ v & 1+v \end{bmatrix}$ & $\begin{bmatrix} 0 & v \\ 1 & 1+v \end{bmatrix}$ \\
\hline
$\begin{bmatrix} 1+v & v \\ 1 & 0 \end{bmatrix}$ & $\begin{bmatrix} 1+v & 1 \\ v & 0 \end{bmatrix}$ \\
\hline \hline
\end{tabular}
\end{table}

\begin{proposition} \label{abk}
For any $k \in \R_2$, $A \in M_2(\R_2)$ is two-sided $k$-orthogonal if and only if $A$ can be expressed in the form  $$A=\left(\begin{matrix} a & b \\ b & a \end{matrix}\right)$$ where $a+b=k$. Consequently, $|O_2(k,\R_2)|=4$. 
\end{proposition}
\pf The proof is tedious but rather straightforward, by using the fact that every element of $\R_2$ is idempotent. \Q{\Box} 
\vskip .1in
\begin{example}
The two-sided $k$-orthogonal semigroups in $M_2(\R_2)$ are $$ O_2(0, \R_2) = \left\{\begin{bmatrix} 0 & 0 \\ 0 & 0 \end{bmatrix},\begin{bmatrix} 1 & 1 \\ 1 & 1 \end{bmatrix}, \begin{bmatrix} v & v \\ v & v \end{bmatrix}, \begin{bmatrix} 1+v & 1+v \\ 1+v & 1+v \end{bmatrix}\right\}$$
$$ O_2(1, \R_2) = \left\{\begin{bmatrix} 1 & 0 \\ 0 & 1 \end{bmatrix},\begin{bmatrix} 0 & 1 \\ 1 & 0 \end{bmatrix}, \begin{bmatrix} v & 1+v \\ 1+v & v \end{bmatrix}, \begin{bmatrix} 1+v & v \\ v & 1+v \end{bmatrix}\right\}$$
$$ O_2(v, \R_2) = \left\{\begin{bmatrix} v & 0 \\ 0 & v \end{bmatrix},\begin{bmatrix} 1 & 1+v \\ 1+v & 1 \end{bmatrix}, \begin{bmatrix} 0 & v \\ v & 0\end{bmatrix}, \begin{bmatrix} 1+v & 1 \\ 1 & 1+v \end{bmatrix}\right\}$$
$$ O_2(1+v, \R_2) = \left\{\begin{bmatrix} 1+v & 0 \\ 0 & 1+v \end{bmatrix},\begin{bmatrix} 1 & v \\ v & 1 \end{bmatrix}, \begin{bmatrix} 0 & 1+v \\ 1+v & 0\end{bmatrix}, \begin{bmatrix} 1+v & 1 \\ 1 & 1+v \end{bmatrix}\right\}$$
These semigroups can actually be generated via $X +kJ_2$, where $X \in O_2(0,\R_2)$ and $J_2$ is the reversal matrix. Note that $O_2(1,\R_2)$ is a Klein-$4$ group.  
\end{example}

In the classical case of a field $F$, a matrix in $M_n(F)$ is left $1$-orthogonal if and only if it is right $1$-orthogonal, as indicated in (\ref{ortho}), that is, $LO_n(1,F)=RO_n(1,F)=O_n(F)$, the group of two-sided orthogonal $n \times n$ matrices over $F$. However still, $LO_n(0,\FF) \ne RO_n(0,F)$ as given in the example below. We extend this result further from fields to rings in Theorem~\ref{group}.
\begin{example} Let $\FF_2$ be the binary field. The one-sided $0$-orthogonal semigroups in $M_2(\FF_2)$ are
\begin{equation} \nonumber LO_2(0, \FF_2) = \left\{\begin{bmatrix} 0 & 0 \\ 0 & 0 \end{bmatrix},\begin{bmatrix} 1 & 1 \\ 1 & 1 \end{bmatrix}, \begin{bmatrix} 0 & 1 \\ 0 & 1 \end{bmatrix}, \begin{bmatrix} 1 & 0 \\ 1 & 0 \end{bmatrix}\right\}\end{equation}
\begin{equation} \nonumber RO_2(0, \FF_2) = \left\{\begin{bmatrix} 0 & 0 \\ 0 & 0 \end{bmatrix},\begin{bmatrix} 1 & 1 \\ 1 & 1 \end{bmatrix}, \begin{bmatrix} 0 & 0 \\ 1 & 1 \end{bmatrix}, \begin{bmatrix} 1 & 1 \\ 0 & 0 \end{bmatrix}\right\}\end{equation}
It follows that $O_2(0, \FF_2) = \left\{\begin{bmatrix} 0 & 0 \\ 0 & 0 \end{bmatrix},\begin{bmatrix} 1 & 1 \\ 1 & 1 \end{bmatrix}\right\}.$
\end{example}
\begin{theorem} \label{group}
For any commutative ring $R$ with unity $1 \neq 0$ and positive integer $n\ge 2$, the multiplicative semigroups $LO_n(1,R)$ and $RO_n(1_{R},R)$ are groups.
\end{theorem}
\pf  Let the set $LO_n(1,R)=\{A\in M_n(R)\;|\;A^TA=I_n \}$. Clearly, $I_n \in LO_n(1,R)$ and is the two-sided identity element, in particular, $I_n$ is the left identity element. Let $A\in LO_n(1,R)$. Since $A^TA=I_n$, then $A^T$ is the left inverse of $A$. Hence $LO_n(1_{R},R)$ contains a left identity element and each element has a left inverse. Therefore, $LO_n(1,R)$ is group. The proof for $RO_n(1,R)$ is analogous. \Q{\Box}

\begin{corollary}
For any commutative ring $R$ with unity $1 \neq 0$ and positive integer $n \ge 2$, we have $LO_n(1,R)=RO_n(1,R) = O_n(R)$.
\end{corollary}
\pf Suppose $A\in LO_n(1,R)$. Then $A^TA=I_n$. Now we take the product $(AA^T)(AA^T)= A(A^TA)A^T = A(I_n)A^T = AA^T$. Since $LO_n(1,R)$ is a group, then by cancellation, $AA^T=I_n$ so that $A\in RO_n(1,R)$. Hence, $LO_n(1,R)\subseteq RO_n(1,R)$. Similarly, $RO_n(1,R)\subseteq LO_n(1,R)$ . The second equality is a mere consequence of (\ref{twosided}). \Q{\Box}
\vskip .1in
The converse of Theorem~\ref{group} is obviously true, as shown below. 
\begin{proposition}
Let $R$ be any commutative ring with unity $1 \neq 0$. If $O_n(k,R)$ is a group, then $k=1$.
\end{proposition}
\pf Since $O_n(k,R)$ is a subgroup of $GL_n(R)$, necessarily the matrix $I_n$, which is in $O_n(1,R)$, is also the identity of $O_n(k,R)$, and we are done. \Q{\Box} 

\subsection{Semigroup isomorphisms}
\label{subsec:iso}
Analogous semigroup isomorphisms for the left [resp. right] $k$-orthogonal semigroups $LO_n(k,R)$ [resp. $RO_n(k,R)$] are deduced and specific examples of these isomorphisms are presented in this section. But initially, we give an easy but useful corollary to Theorem~\ref{han2006}. On the side, we give a product formula for the cardinality of $O_n(R)$.  

\begin{corollary}
For any positive integers $m,n\geq 2$, if $R$ is a finite semi-local ring with maximal ideals $I_1,I_2,\ldots,I_m$, then \begin{equation} \label{orthoprod} O_n(R)\cong O_n(\FF_{q_1})\times O_n(\FF_{q_2})\times \ldots \times O_n(\FF_{q_m}) \end{equation} where $\FF_{q_j}$ are Galois fields of order $q_j$, $j=1,2,\ldots,m$. 
\end{corollary}

\pf In Theorem~\ref{han2006}, $GL_n(R)$ is isomorphic to the product $GL_n(R/I_1)\times GL_n(R/I_2)\times...\times GL_n(R/I_m)$. Since $R$ is finite and $I_j$ is maximal, then $R/I_j$ is a Galois field $\FF_{q_j},$ for some $q_j$ a power of a prime. That is, $$ GL_n(R)\cong GL_n(\FF_{q_1})\times GL_n(\FF_{q_2})\times \ldots \times GL_n(\FF_{q_m})$$ Now, $O_n(R)$ is a subgroup of $GL_n(R)$, and $O_n(\FF_{q_j})$ is a subgroup of $GL_n(\FF_{q_j})$, for $j=1,2,\ldots,m$. By extension and the correspondence theorem, the result (\ref{orthoprod}) follows. \Q{\Box} 
\vskip .1in
Consequently the number of $n\times n$ orthogonal matrices over $R$ is given by \begin{equation} \label{orthocard} |O_n(R)|=|O_n(\FF_{q_1})|\cdot|O_n(\FF_{q_2})|\cdot .... \cdot|O_n(\FF_{q_m})| \end{equation}
where each factor in the product is computed using the formula from \cite{macw}.

\begin{example} \label{exa2} We have $GL_2(\R_2) \cong GL_2(\FF_2) \times GL_2(\FF_2)$ Using the formula in (\ref{general}) we get $|GL_2(\R_2)|=6 \cdot 6 = 36$. Further, $O_2(1,\R_2) \cong O_2(\FF_2) \times O_2(\FF_2)$, which is a Klein-$4$ group. Since $|O_2(\FF_2)|=2$ from \cite{macw}, therefore $|O_2(1,\R_2)|=4$, as we already know. 
\end{example}

\begin{example} We have $O_2(1,\ZZ_6) \cong O_2(\FF_2) \times O_2(\FF_3)$ Now, $|O_2(\FF_2)|=2$ and $|O_2(\FF_3)|=8$ from the formulas in \cite{macw}. Hence, $\abs{O_2(1,\ZZ_6)}=16$ as earlier mentioned in Example~\ref{exa16}. In fact, $O_2(1,\ZZ_6) \cong \ZZ_2 \times D_4$, where $D_4$ is the dihedral group of degree $4$.
\end{example}

\begin{corollary}
Let $R$ be a finite semi-local ring and $k\neq 1$ be an idempotent element of $R$. For positive integers $m,n\geq2$, we have \begin{equation} \label{lorthoprod} LO_n(k,R)\cong LO_n(a_1,\FF_{q_1})\times LO_n(a_2,\FF_{q_2})\times \ldots \times LO_n(a_m,\FF_{q_m}) \end{equation}
where the $q_i$'s are powers of primes, and the $a_i$'s are determined by the mapping from $R$ to $\FF_{q_1}\times\FF_{q_2}\times \ldots \times\FF_{q_m}$. Analogously for $RO_n(k,R)$.
\end{corollary}

\pf We consider the isomorphism $\theta$ in Theorem~\ref{han2006} as an isomorphism of multiplicative semigroups and follow the same reasoning in the previous corollary. Since $R$ is finite and $I_j$ is maximal, then $R/I_j$ is a Galois field $\FF_{q_j},$ for some $q_j$ a power of a prime. Now, $LO_n(k,R)$ is a subsemigroup of $M_n(R)$, and $LO_n(a_j,\FF_{q_j})$ is a subsemigroup of $M_n(\FF_{q_j})$ for $j=1,\dots,m$. By extension and the correspondence theorem for semigroups, the result (\ref{lorthoprod}) follows. \Q{\Box} 
\vskip .1in
Consequently the number of left $k$-orthogonal matrices over $R$ is given by \begin{equation} |LO_n(R)|=|LO_n(a_1, \FF_{q_1})|\cdot|LO_n(a_2, \FF_{q_2})|\cdot .... \cdot|LO_n(a_m, \FF_{q_m})| \end{equation}

\begin{example}
The Boolean ring $\R_2=\FF_2+v\FF_2$, $v^2=v$ is isomorphic to the product $\FF_2 \times \FF_2$ via the map $a+vb \mapsto (a+b,a)$. Hence, $$LO_n(a+vb,\R_2) \cong LO_n(a+b,\FF_2)\times LO_n(a,\FF_2).$$ In particular, we have the following isomorphisms.

$$LO_n(0,\FF_2+v\FF_2)\cong LO_n(0,\FF_2)\times LO_n(0,\FF_2)$$ 
$$LO_n(v,\FF_2+v\FF_2)\cong O_n(\FF_2)\times LO_n(0,\FF_2)$$ 
$$LO_n(1+v,\FF_2+v\FF_2)\cong LO_n(0,\FF_2)\times O_n(\FF_2)$$
\end{example}

\begin{example} \label{six}
The integer ring $\ZZ_6$ is isomorphic to $\FF_2\times\FF_3$ via the map $x \mapsto (x\mod 2, x\mod3)$. It follows that

$$LO_n(0,\ZZ_6)\cong LO_n(0,\FF_2)\times LO_n(0,\FF_3)$$
$$LO_n(3,\ZZ_6)\cong O_n(\FF_2)\times LO_n(0,\FF_3)$$
$$LO_n(4,\ZZ_6)\cong LO_n(0,\FF_2)\times O_n(\FF_3) $$
\end{example}

\begin{example} \label{f2r}
The set $\FF_{2^r}+v\FF_{2^r}=\{a+vb\;|\;a,b\in\FF_{2^r},\;v^2=v\}$ is a commutative semi-local ring with unity under addition and multiplication modulo $2$. This ring has exactly two proper nontrivial ideals $\langle v\rangle=\{xv\;|\;x\in\FF_{2^r}\}$ and $\langle1+v\rangle=\{x(1+v)\;|\;x\in\FF_{2^r}\}$. These ideals are both maximal and $|\langle v\rangle|=|\langle1+v\rangle|=2^r$. There are $2(2^r-1)$ zero divisors and $(2^r-1)^2$ units. Now, $\FF_{2^r}+v\FF_{2^r}\cong \FF_{2^r}\times\FF_{2^r}$ via the map $a+vb \mapsto (a+b,a)$ so that $LO_n(a+vb,\FF_{2^r}+v\FF_{2^r})\cong LO_n(a+b,\FF_{2^r})\times LO_n(a,\FF_{2^r}).$
\end{example}

\begin{example} \label{fpr}
For $p$ prime, $p\neq 2$, the set $\FF_{p^r}+v\FF_{p^r}=\{a+vb\;|\;a,b\in\FF_{p^r},\;v^2=1\}$ is a commutative semi-local ring with unity under addition and multiplication modulo $p$. It has exactly two maximal ideals $\langle1+v\rangle=\{x(1+v)\;|\;x\in\FF_{p^r}\}$ and $\langle1-v\rangle=\{x(1-v)\;|\;x\in\FF_{p^r}\}$ with $|\langle1+v\rangle|=|\langle1-v\rangle|=p^r$. There are $2(p^r-1)$ zero divisors and $(p^r-1)^2$ units. Now, $\FF_{p^r}+v\FF_{p^r}\cong \FF_{p^r}\times\FF_{p^r}$ via the map $a+vb \mapsto (a-b,a+b)$ so that $LO_n(a+vb,\mathbb{F}_{p^r}+v\mathbb{F}_{p^r})\cong LO_n(a+b,\mathbb{F}_{p^r})\times LO_n(a-b,\mathbb{F}_{p^r}).$
\end{example}

We summarize in the tables that follow the cardinalities of the finite semigroups of left and right $k$-orthogonal matrices over $\R_2$ and $\ZZ_6$. MAGMA${}^{\circledR}$ programs are written to create the semi-local ring and identify its idempotents, then for $n=2,3$ and for each idempotent $k$, elements of the matrix ring are exhaustively tested and counted for the semigroup. The resulting cardinalities are compared with the product formulas derived above. With slight adjustments, the routines can work for other rings as well.  

\begin{table}[h]
\centering
\caption{$k$-Orthogonal $2\times2$ Matrices Over $\R_2$}
\begin{tabular}{c|c|c|c}
\hline
\hline
$k$ & $LO_2(k,\R_2)=RO_2(k,\R_2)$ & $O_2(k,\R_2)$ & $LO_2(k,\R_2)-O_2(k,\R_2)$ \\
\hline
$0$ & $16$& $4$ & $12$ \\
$1$ & $4$ & $4$ & $0$  \\
$v$ & $8$ & $4$ & $4$  \\
$1+v$ & $8$ & $4$ & $4$  \\
\hline
\hline
\end{tabular}
\end{table}

\begin{table}[h]
\centering
\caption{$k$-Orthogonal $3\times3$ Matrices Over $\R_2$}
\begin{tabular}{c|c|c|c}
\hline
\hline
$k$ & $LO_3(k,\R_2)=RO_3(k,\R_2)$ & $O_3(k,\R_2)$ & $LO_3(k,\R_2)-O_3(k,\R_2)$\\
\hline
$0$ & $484$ & $100$ & $384$ \\
$1$ & $36$  & $36$ & $0$  \\
$v$ & $132$ & $60$ & $72$  \\
$1+v$ & $132$ & $60$ & $72$ \\
\hline
\hline
\end{tabular}
\end{table}

\begin{table}[h]
\centering
\caption{$k$-Orthogonal $2\times2$ Matrices Over $\ZZ_6$}
\begin{tabular}{c|c|c|c}
\hline
\hline
$\; \; \; k \; \; \; $ & $LO_2(k,\ZZ_6)=RO_2(k,\ZZ_6)$ & $O_2(k,\ZZ_6)$ & $LO_2(k,\ZZ_6)-O_2(k,\ZZ_6)$  \\
\hline
$  0  $ & $4$ & $2$ & $2$  \\
$  1  $ & $16$ & $16$ & $0$  \\
$  3  $ & $2$ & $2$ & $0$  \\
$  4  $ & $32$ & $16$ & $16$  \\
\hline
\hline
\end{tabular}
\end{table}

\begin{table}[ht]
\centering
\caption{$k$-Orthogonal $3\times3$ Matrices Over $\ZZ_6$}
\begin{tabular}{c|c|c|c}
\hline
\hline
$\; \; \; k \; \; \;$ & $LO_3(k,\ZZ_6)=RO_3(k,\ZZ_6)$ & $O_3(k,\ZZ_6)$ & $LO_3(k,\ZZ_6)-O_3(k,\ZZ_6)$ \\
\hline
$  0  $ & $2310$ & $330$ & $1980$ \\
$  1  $ & $288$  & $288$ & $0$ \\
$  3  $ & $630$  & $198$ & $432$ \\
$  4  $ & $1056$ & $480$ & $576$ \\
\hline
\hline
\end{tabular}
\end{table}

\subsection{Codes from $k$-orthogonal matrices}

We study how $(-1)$-orthogonal matrices and $0$-orthogonal matrices can give rise to leading-systematic self-dual or weakly self-dual linear codes. The construction of these classes of codes is illustrated in some examples.

In this section, we let $R$ to be a finite commutative ring with unity. Following the discussion in the previous section, we use the terminology in \cite{mas1} to say that a matrix $A\in M_n(R)$ is left [resp. right] antiorthogonal if and only if $A$ is left [resp. right] $(-1)$-orthogonal. If $A$ is both left antiorthogonal and right antiorthogonal, then $A$ is said to be an {\it antiorthogonal} matrix. However, the following proposition says that one-sided antiorthogonal matrices over $R$ coincide. 

\begin{proposition}
A left antiorthogonal matrix is a right antiorthogonal matrix, and conversely.
\end{proposition} 
\pf Suppose $AA^T=-I_n$. Then $\det A$ is a unit in $R$ so that $A \in GL_n(R)$ in which left and right inverses coincide. The result follows.\Q{\Box} 
\vskip .1in
The matrix in Example~\ref{firstex} is an antiorthogonal matrix. Clearly, an antiorthogonal matrix is orthogonal if and only if the characteristic of $R$ is $2$. Consequently, Proposition~$1$ of \cite{mas1} can be stated more generally as follows.  
\begin{proposition}
Let $R$ be a finite commutative ring with unity. A rate-$k/n$ leading-systematic linear code over $R$ is self-dual if and only if, in its generator matrix $G=[I_k : A]$, the matrix $A$ is antiorthogonal.
\end{proposition} 
It should be noted that, in this case, $n=2k$ so that $A$ is also a $k \times k$ matrix. 
\vskip .1in
The MAGMA${}^{\circledR}$ routines that we have developed help us to find the one-sided $k$-orthogonal matrices, and in particular, the antiorthogonal matrices which subsequently give the self-dual codes, as illustrated in the example below. The self-duality is checked further in MAGMA${}^{\circledR}$.   

\begin{example}
Consider the two-sided $5$-orthogonal matrix $A=\begin{bmatrix} 4 & 5 \\ 1 & 4 \end{bmatrix}\in M_2(\ZZ_6)$. Then form the $2 \times 4$ generator matrix $$G=\begin{bmatrix} I_2 & A \end{bmatrix}=\begin{bmatrix} 1 & 0 & 4 & 5 \\ 0 & 1 & 1 & 4 \end{bmatrix}.$$
The leading-systematic code over $\ZZ_6$ generated by $G$ is self-dual.
\end{example}

\begin{example} \label{octacode}
The rate-$4/8$ octacode $\mathcal O_8$ over $\ZZ_4$ is a self-dual leading-systematic linear code with generator matrix $G=[I_4 : A]$ where $$A=\begin{bmatrix} 3 & 1 & 2 & 1 \\ 1 & 2 & 3 & 1 \\ 3 & 3 & 3 & 2 \\ 2 & 3 & 1 & 1 \end{bmatrix}$$ is a two-sided $3$-orthogonal matrix. The Lee distance is $6$ and the Hamming distance is $4$. 
\end{example}

The antiorthogonality condition can be extended to non-square matrices. A $k\times m$ matrix $A$ is said to be {\it right row-antiorthogonal} if $AA^T=-I_k$. Deleting rows of an antiorthogonal matrix will result to a right row-antiorthogonal matrix, but not every right row antiorthogonal matrix can be constructed. We claim an extension of Proposition~$2$ in \cite{mas1} as follows. 

\begin{proposition}
Let $R$ be a finite commutative ring with unity. A leading-systematic linear code over $R$ is weakly self-dual if and only if, in its generator matrix $G=\begin{bmatrix} I_k & A \end{bmatrix}$, the matrix $A$ is right row-antiorthogonal.
\end{proposition}

\begin{example}
With the matrix $A$ of the octacode $\mathcal O_8$ in Example~\ref{octacode}, delete the fourth row to get $B=\begin{bmatrix} 3 & 1 & 2 & 1 \\ 1 & 2 & 3 & 1 \\ 3 & 3 & 3 & 2 \end{bmatrix}$. It can be verified that $B$ is a right row-antiorthogonal matrix. We then construct the leading-systematic rate-$3/7$ $\ZZ_4$-linear code using the generator matrix $$G=\begin{bmatrix} I_3 & B \end{bmatrix}=
\begin{bmatrix} 1 & 0 & 0 & 3 & 1 & 2 & 1 \\ 0 & 1 & 0 & 1 & 2 & 3 & 1 \\ 0 & 0 & 1 & 3 & 3 & 3 & 2 \end{bmatrix}.$$ The code generated by $G$ is weakly self-dual. The Lee distance is $6$ and the Hamming distance is $4$. 
\end{example}

A matrix $A\in M_n(R)$ is said to be left [resp. right] self-orthogonal if and only if $A$ is left [resp. right] $0$-orthogonal. If $A$ is both left self-orthogonal and right self-orthogonal, then the matrix $A$ is simply called {\it self-orthogonal}. The matrix in Example~\ref{selforth} is a self-orthogonal matrix. Unlike the case of antiorthogonal matrices, left self-orthogonal matrices and right self-orthogonal matrices do not necessarily coincide. 

\begin{example}
The matrix in  $M_3(\ZZ_6)$ given by $$A=\begin{bmatrix} 0 & 3 & 3 \\ 4 & 2 & 4 \\ 2 & 1 & 5 \end{bmatrix}$$ is right self-orthogonal but not left self-orthogonal. We adopt the analogous idea of a row right self-orthogonal matrix. Deleting any row of $A$, say, the third row, we get the $2 \times 3$ matrix $G=\begin{bmatrix}  0 & 3 & 3 \\ 4 & 2 & 4 \end{bmatrix}$, which is a row right self-orthogonal matrix. The matrix $G$ generates a weakly self-dual code over $\ZZ_6$.   
\end{example}
This example is generalized in the proposition below.
\begin{proposition}
Let $R$ be a finite commutative ring with unity. A linear code over $R$ with generator matrix $G$ is weakly self-dual if and only if $G$ is row right self-orthogonal.
\end{proposition}

\section{Conclusion and recommendations}
\label{sect:recom}
For further research it is interesting to derive explicit formulas for the cardinality of the finite one-sided $k$-orthogonal semigroups. It is also worthy to consider applying the results of this paper to other classes of finite semi-local rings with multiple idempotents such as those given in Examples \ref{f2r} and \ref{fpr}. More specific and relevant structural properties of the matrix semigroups can be explored. 

We have also proved that there exist no $3 \times 3$ antiorthogonal matrices over the integer ring $\ZZ_6$. This is accomplished by using the map from $\ZZ_6$ onto $\FF_2 \times \FF_3$ in Example~\ref{six} to show that indeed $-1$ is a non-square in $\ZZ_6$, and by \cite{mas2} that, for any odd integer $n$, there exist no $n \times n$ antiorthogonal matrices over $\FF_3$. This specific result is likewise verified through exhaustive search. The reader is encouraged to determine the conditions for the existence of antiorthogonal matrices over finite semi-local rings.   

Finally, the connections of the above defined matrices to the characterization of LCD codes over finite commutative rings with unity can be studied.   

\section{Acknowledgement}
The first author gratefully acknowledges the University of the Philippines for the One U.P. Professorial Chair Award in Mathematics (Coding Theory) for Outstanding Research and Teaching in U.P. Los Ba\~{n}os from 2016 to 2021.

\end{document}